\documentclass{jkas}

\def\year{2014} 
\def\month{March} 
\def\volume{00} 
\def\issue{0} 
\def\beginpage{1} 
\def\endpage{99} 
\def\received{---} 
\def\accepted{---} 

\date{Received \received ; accepted \accepted}

\title{
A New Method to Calibrate the Stellar Color/Surface-Brightness Relation
}

\author{Andrew Gould}

\affil{Department of Astronomy Ohio State University,
140 W.\ 18th Ave., Columbus, OH 43210, USA 
\email{gould@astronomy.ohio-state.edu }}

\def\corrauthor{
Andrew Gould
}

\def\runningauthor{
Andrew Gould
}

\def\runningtitle{
Stellar Surface Brightness Calibration
}

\def\keywords{
astrometry -- gravitational microlensing -- stars: fundamental parameters (radii)
}

\def\abstracttext{
I show that the standard microlensing technique to measure the angular
radius of a star using color/surface-brightness relations can be inverted,
via late-time proper motion measurements,
to calibrate these relations.  The method is especially useful for
very metal-rich stars because such stars are in short supply in the solar
neighborhood where other methods are most effective, but very abundant
in Galactic bulge microlensing fields.  I provide a list of eight
spectroscopically identified high-metallicity bulge stars with
the requisite finite-source effects, seven of which will be suitable
calibrators when the Giant Magellan Telescope comes on line.  Many
more such sources can be extracted from current and future microlensing
surveys.
}

\newcommand{\bdv}[1]{\mbox{\boldmath$#1$}}

\def\au{{\rm AU}} 
 
\def\bv{{\bf v}}

\def\kms{{\rm km}\,{\rm s}^{-1}}
\def\masyr{{\rm mas}\,{\rm yr}^{-1}}
\def\muas{{\mu\rm as}}

\def\rel{{\rm rel}}

\def\eff{{\rm eff}}
\def\hel{{\rm hel}}
\def\geo{{\rm geo}}
\def\feh{{\rm [Fe/H]}}
\def\e{{\rm E}}
\def\bpi{{\bdv\pi}}
\def\bmu{{\bdv\mu}}

\def\la{{<\atop \sim}}
\def\ga{{>\atop \sim}}
\def\apj{{ApJ}}
\def\apjl{{ApJL}}
\def\aap{{A\&A}}
\def\pasp{{PASP}}
\def\mnras{{MNRAS}}

\begin{document}
\jkashead 


\section{{Introduction}
\label{sec:intro}}

Stellar radii $R$ are fundamental parameters but are directly measured
for a tiny fraction of all stars.  They can, however, be accurately
estimated from $R=d\theta_*$ if the distance $d$ is known and the
angular radius $\theta_*$ can be determined using color/surface-brightness
relations.  For example, \citet{kervella04} give
\begin{equation}
K_0 + 5\log(\theta_*/{\rm mas}) = 0.377(V-K)_0 + 1.080
\label{eqn:kerv}
\end{equation}
for dwarfs and subgiants.  The lhs is (a logarithm of) the
mean surface brightness of the star, while the rhs is proportional
to the color.

\citet{kervella04} argue that there is no evidence
for a dependence of this relation on metallicity over the
range $-0.5<\feh<+0.5$.  However, while their Figure~5  indeed provides
no evidence for a metallicity dependence, neither does it provide strong
evidence against this hypothesis.  Such an effect would be expected
to appear most strongly in cool metal-rich stars, for which
line blanketing in the $V$ band (and other blue bands) is severe.
That is, Equation~(\ref{eqn:kerv}) implicitly uses $(V-K)$ as a temperature
indicator, but this indicator will be faulty if the $V$ flux is suppressed
by cool metals in the atmosphere.  In the \citet{kervella04} sample
of dwarfs and subgiants, it is mainly the subgiants that are cool,
and none of these is more metal-rich than $\feh=+0.25$.
It would be useful
to carry out the calibration at the high-metallicity end by
including additional stars with precise $\theta_*$ measurements
for metal-rich stars, and in particular, subgiants and giants.

\begin{figure}
\centering
\includegraphics[width=80mm]{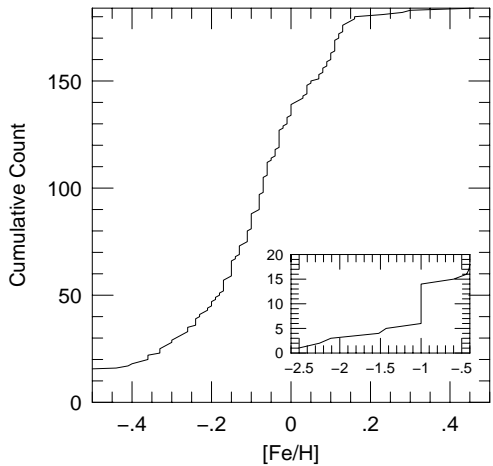}
\caption{Cumulative distribution of 
``giant stars'' ($R>6R_\odot$ and $V-K>1.5$)
with angular diameter measurements from \citet{charm2} and metallicities
from \citet{hipext}.  The interval $-0.3<{\rm [Fe/H]}<0.2$ is well sampled
but there are only four stars ${\rm [Fe/H]}>0.16$.}
\label{fig:fehcum}
\end{figure}

To assess the difficulty of doing so from a local sample, I
search the CHARM2 catalog \citep{charm2} for 
`red giants'', defined as having radii $R>6 R_\odot$
and color $V-K>1.5$, with angular diameter measurements from 
lunar occultations, long-baseline interferometry, fine-guidance sensor,
speckle interferometry, and aperture masking (with the great majority
coming from the first two).  I find 372 such stars of which 184 have
metallicities listed in \citet{hipext}. Figure~\ref{fig:fehcum} shows
the cumulative distribution of these metallicities.

Note that there are only four stars with 
${\rm [Fe/H]}>0.16$: HIP (22729, 88567, 85617, 100345) at 
[Fe/H] (0.23, 0.28, 0.30, 0.46).  Furthermore, the first of these
has an unusably poor (23\%) $\theta_*$ measurement, and the next two
are pulsating variables.  Thus, the high-metallicity end of the
distribution is extremely poorly probed.

\section{{Microlensing and Angular Radii}
\label{sec:ulens}}

Microlensing studies make routine use of color/surface-brightness
relations to estimate $\theta_*$, which they use as an intermediate
step to determine the ``Einstein radius'' $\theta_\e$, via the
relation $\theta_\e =\theta_*/\rho$.  Here, $\rho$ is a parameter
that is returned by microlensing models, essentially whenever
the source is observed to cross a caustic.  That is, the duration
of the source-caustic-crossing time relative to the Einstein timescale $t_\e$
of the event as a whole allows to determine $\rho=\theta_*/\theta_\e$.

As discussed by \citet{ob03262}, the mathematical model of the microlensing
event yields the instrumental magnitudes of the source (free of blending)
in several bands, usually including $V$ and $I$.  One finds the offset
of these values $\Delta ((V-I),I)$
from the red clump (RC) centroid, using the same instrumental
photometry.  The dereddened color of the clump is known to be 
$(V-I)_{0,\rm cl}=1.06$ from the work of \citet{bensby13}, 
while the dereddened magnitude $I_{0,\rm cl}$ is known as a function
of field position from the study by \citet{nataf13}.  These yield
$(V-I,I)_0 = (V-I,I)_{0,\rm cl} + \Delta ((V-I),I)$.  Then this
$V/I$ photometry is converted to $V/K$ using the empirical color-color
relations of \citet{bb88}.  The $(V-K)_0$ color is used to estimate
the $K$-band surface brightness employing the empirical color/surface-brightness
relations of \citet{kervella04}, and finally this is combined with $K_0$
to determine the source angular radius, $\theta_*$.  

There are two
principal sources of uncertainty in this estimate.  First, the dereddened
color is determined only to about $\sigma(V-I)_0\sim 0.05$ mag.  This
uncertainty is known because the color-estimation procedure has been applied by
\citet{bensby13} to a sample of about 50 dwarfs and subgiants with
high-resolution spectra (taken when the source was highly magnified
by microlensing).  Then the $(V-I)_0$ colors were predicted from
models based on spectral classification and compared to those
determined by the microlensing method.  For relatively blue stars
near the turnoff, the scatter is about 0.06 mag, of which some contribution
is due to the uncertainty in the spectroscopic temperature, implying
that the intrinsic scatter in the microlensing method is 0.05 mag
(or possibly less, if there are other unrecognized errors in the
spectroscopic determinations).  Redder microlensed stars show greater
scatter but \citet{bensby13} argue that this is due to uncertainty
in the spectroscopic models of these stars.  Note that \citet{bensby13}
determine the color of the RC by choosing the value that minimizes
this scatter.

Second, there is typically a 0.1 mag uncertainty in estimating the $I$-band
magnitude of the
RC centroid.  These two errors combined yield a 7\% error in $\theta_*$.
No account is usually taken of errors in the overall distance scale
(i.e., $R_0$) derived by \citet{nataf13} nor in the 
color/surface-brightness relations derived by \citet{kervella04}, since
these are deemed small compared to the dominant errors.

\section{{Using Microlensing to Calibrate the Relation}
\label{sec:calib}}

The parameter $\rho$ is not always measured in microlensing events,
but when it is, the above process can in principle
be inverted to {\it measure} $\theta_*$ and so measure the surface
brightness.  The events for which $\rho$
is measurable are also those for which it is easiest to get a
spectrum because the sources are either big (so bright) and
so more likely to transit a caustic, or because the sources are
highly magnified, which also makes caustic crossings more likely.
Hence, these are the same stars for which it is most feasible
to measure a metallicity.

The basic method is simply to wait for the source and lens to
separate enough to be clearly resolved in high-resolution (e.g.,
adaptive optics -- AO) images.   This automatically yields a measurement
of their separation $\Delta\theta$, and hence their heliocentric
proper motion
\begin{equation}
\mu_\hel = {\Delta\theta\over\Delta t},
\label{eqn:helio}
\end{equation}
where $\Delta t$ is the time interval from the peak of the event
to the time of observation.  Equation~(\ref{eqn:helio}) makes
two approximations.  First, it assumes that the lens and source
were perfectly aligned at the peak of the event.  For typical events,
they are misaligned by $\la 100\,\muas$, whereas $\Delta\theta$
will typically be many tens of mas (to enable separate resolution).
Hence, the error induced by this approximation is usually negligible
and, in any case, quantifiable.  Second it ignores the lens-source
parallactic motion due to their relative parallax $\pi_\rel$.
There are three points about this approximation.  First, it is
also typically the case that $\pi_\rel\la 100\,\muas$, so this
effect is similarly small.  Second, if one were really worried about
this effect, one could make the AO measurement at the same time of
year as the peak of the event.  Finally, this effect is generally smaller
than one I discuss below that is also directly proportional to $\pi_\rel$.
Hence, for purposes of discussion, I simply use Equation~(\ref{eqn:helio})
as is.

Next, if $\rho$ (and $t_\e$) are measured, then their product
$t_* \equiv \rho t_\e$ is also measured.  Actually, $t_*$
is typically measured to much higher precision than either
$\rho$ or $t_\e$ separately,  particularly in the high-magnification
events \citep{mb11293}.
This is because $t_*$ reflects the caustic crossing time,
which is a direct observable, whereas $\rho$ and $t_\e$ are covariant
with many other parameters, including each other.

Naively, then, we have $\theta_* = \mu t_*$, and we appear to be done.  
Unfortunately, the caustic crossing time is measured in the frame
of Earth at the peak of the event, while $\mu_\hel$ is measured
in the Sun frame, as described above.  Hence, the appropriate equation is
\begin{equation}
\theta_* = \mu_\geo t_*,
\label{eqn:thetastar1}
\end{equation}
where \citep{mb08310}
\begin{equation}
\bmu_\geo = \bmu_\hel - \bmu_\oplus\pi_\rel;
\qquad
\bmu_\oplus \equiv {\bv_{\oplus,\perp}\over \au},
\label{eqn:thetastar2}
\end{equation}
and $\bv_{\oplus,\perp}$ is the transverse velocity of Earth
in the frame of the Sun at the peak of the event.

\subsection{{Uncertainty of $\theta_*$ Measurement}
\label{sec:uncert}}

Of course, $\bmu_\oplus$ is known with extremely high precision,
but $\pi_\rel$ may not be known very well, and this can lead to
significant uncertainty in $\theta_*$ even if $\mu_\hel$ and $t_*$
are well measured.  To gain a sense of this, I note that the great
majority of usable microlensing events peak within 2 months of opposition and
also that it is only the component of $\bmu_\oplus$ that is
aligned with $\bmu_\hel$ that plays a significant role in 
Equation (\ref{eqn:thetastar2}).  Therefore,
I adopt $\bmu_\oplus\cdot \bmu_\hel/\mu_\hel=3.5\,{\rm yr}^{-1}$
as a typical value.  In this case, an uncertainty in $\pi_\rel$ of
$100\,\muas$ leads to an uncertainty in $\mu_\hel$ of $0.35\,\masyr$,
which should be compared to typical values of $\mu_\hel$ of
$4\,\masyr$ and $7\,\masyr$ for bulge and disk lenses respectively.
Hence, uncertainties in $\pi_\rel$ must be minimized.

There are two main routes to doing so.  First, since the lens is
resolved, the same high-resolution images that measure its
position can also be used to measure its color and magnitude,
and from this one can estimate a photometric distance.  
Given the fact that the distribution of the dust along the
line of sight is not very well known and that the photometry
is likely to be mostly in the infrared, such an estimate, by itself,
would be fairly crude.  However, there is an additional constraint
on the lens mass $M$ and the relative parallax $\pi_\rel$
from (e.g., \citealt{gould00})
\begin{equation}
M\pi_\rel = {\theta_\e^2\over\kappa} = {(\mu_\geo t_\e)^2\over\kappa};
\qquad
\kappa\equiv {4 G\over c^2\,\au}\simeq 8.1{{\rm mas}\over M_\odot}.
\label{eqn:thetae}
\end{equation}

The combination of photometric and $\theta_\e$ constraints can be
extremely powerful.  For example, in a majority of cases, the lens
will be in the Galactic bulge.  It will therefore be behind essentially
all the dust, allowing precise dereddening of the photometry.
Moreover, once constrained to being in the bulge, even very large
{\it relative} changes in $\pi_\rel=\pi_l - \pi_s$ lead to small changes
in distance, so that both the absolute magnitude and color can be
estimated quite precisely.  Then, even with considerable uncertainty
in the stellar absolute-magnitude/mass relation, $\pi_\rel$ can be estimated
very precisely (e.g., \citealt{mb11293B}).  As an example, consider an M dwarf
lens $M=0.5\,M_\odot$, with $\pi_s=110\,\muas$ and $\pi_l=140\,\muas$,
so $\theta_\e = 0.35\,$mas.  Now, $\theta_\e$ will be known with
few percent precision, so if
$M$ were estimated photometrically to 20\%, then Equation~(\ref{eqn:thetae}) 
would lead
to a $\sim 30\%$ error in $\pi_\rel$.  However, since $\pi_\rel$ is
only $30\,\muas$, this would propagate to only a $\sim 1\%$ error
in $\theta_*$.

The real difficulties posed by uncertainty in $\pi_\rel$ come for
disk lenses, which are the minority.  For these, the extinction is
uncertain while modest fractional errors in the lens-mass estimate,
(leading to modest fractional errors in $\pi_\rel$) still yield
relatively large absolute errors in $\pi_\rel$ (and so $\theta_*$) simply
because $\pi_\rel$ is itself relatively large.  However, for disk
lenses there is often another source of information: the microlens
parallax vector $\bpi_\e$,
\begin{equation}
\bpi_\e\equiv {\pi_\rel\over\theta_\e}{\bmu\over\mu}.
\label{eqn:pie}
\end{equation}
This quantity parameterizes the lens-source displacement due to reflex
motion of Earth (though $\pi_\rel$) scaled by the Einstein radius ($\theta_\e$),
and is therefore measurable from the resulting distortions of the lightcurve.
It is a vector because these distortions depend on the direction
of lens-source relative motion ($\bmu$) relative to the ecliptic.  
Because microlensing events are
typically short compared to Earth's orbital time, these effects are
not usually large.  However, disk lenses are an important exception
because for them $\pi_\rel$ (and so $\pi_\e$) can be big.  In particular,
$\pi_{\e,\parallel}\equiv {\bf \hat n}_a\cdot\bpi_\e$, the component of
$\bpi_\e$  parallel to Earth's instantaneous direction of acceleration
projected on the sky, ${\bf \hat n}_a$, is usually much better measured
than $\pi_{\e,\perp}$ \citep{gmb94,smp,gould04}.  
This is because lens-source motion
in this direction leads to an asymmetric distortion in the lightcurve,
which is not easily confused with other microlensing effects.
Note from Equation~(\ref{eqn:pie}) that $\bpi_{\e,\hel}$ and $\bpi_{\e,\geo}$ 
have the same amplitude but different directions.  It is actually
$\pi_{\e,\parallel,\geo}$ that is well-measured in microlensing events.

The first point is that if $\bpi_{\e,\geo}$ is well measured, then one
can determine the projected velocity in the geocentric frame
${\tilde \bv}_\geo = \bpi_{\e,\geo}(\au/\pi_\e^2 t_\e)$ and so solve for
it in the heliocentric
frame ${\tilde \bv}_\hel = {\tilde \bv}_\geo + \bv_{\oplus,\perp}$.  Then,
since $\mu/\tilde v=\pi_\rel/\au$, one can solve directly
for $\bmu_\geo$ and so $\theta_* = \mu_\geo t_*$,
\begin{equation}
\bmu_\geo = \bmu_\hel - \mu_\hel
\bigg| {\bpi_{\e,\geo}\over \pi_\e^2 t_\e} + 
\bmu_\oplus\bigg|^{-1}\bmu_\oplus
\label{eqn:direct}
\end{equation}

To understand the role of microlens parallax measurements more generally, I
write $\bmu_\geo$ in terms of observables.
\begin{equation}
\bmu_\geo= 
{\bmu_\hel} - \theta_\e\pi_\e \bmu_\oplus =
{\bmu_\hel} - \mu_\geo t_\e\pi_\e\bmu_\oplus
\label{eqn:thetastarobs}
\end{equation}
where I have written $\theta_\e = \mu_\geo t_\e$ in the final step,
since $t_\e$ is evaluated during the event, i.e., in the geocentric frame.

As pointed out by \citet{ghosh04}, a good measurement of $\pi_{\e,\parallel,\geo}$
can yield a full measurement of $\bpi_{\e,\geo}$ if one can extract the
direction of $\bpi_{\e,\geo}$ from late time astrometry of the lens and
source after they have moved apart sufficiently to be separately resolved.
However, what is actually measured from such data is $\bmu_\hel$,
and so the direction of $\bpi_{\e,\hel}$, not of $\bpi_{\e,\geo}$.  This implies
that such solutions must be sought self-consistently.  Depending on the
angle between $\bmu_\hel$ and $\bmu_\oplus$, this may be more
or less difficult.  Similarly, in the second form of 
Equation~(\ref{eqn:thetastarobs}), $\mu_\geo$ must also be solved
self-consistently.  Thus, the viability of this approach for disk
lenses must be evaluated on a case by case basis, depending on the
magnitude of the errors in the microlensing parallax vector and in the
value of $\bmu_\oplus\cdot \bmu_\hel/\mu_\hel$ which determines
the fractional amplitude of the correction.

\subsection{{Uncertainty of Surface Brightness Measurement}
\label{sec:uncert2}}

The other elements going into the color/surface-brightness calibration
are the dereddened flux and color measurements $K_0$ and $(V-K)_0$.
For a large fraction of current microlensing events, observations
are routinely carried out in $V$, $I$, and $H$ using the ANDICAM
camera on the 1.3m CTIO-SMARTS telescope \citep{depoy03}, which
employs an optical/infrared dichroic.  This permits very precise
measurement of the (reddened) source flux in these bands for highly
magnified targets because the fractional photometric errors are
small for bright targets and because the magnifications are known
from the microlensing model.  Usually there is a small correction
from $H$ to $K$, which can be evaluated either using the spectral
type (see below) or the late-time astrometric/photometric measurements
when the source and lens are separated.  The problem is to convert these
measurements of $V$ and $K$ to $K_0$ and $(V-K)_0$.  

The method would
be the same as in current microlensing studies (offset from the RC)
except that it would be carried out using a $[(V-K),K]$ color-magnitude
diagram, rather than $[(V-I),I]$.  These color errors are unknown 
at the present time but
for sake of discussion I assume that they are similar to the current
$(V-I)$ errors (particularly taking account of the fact that the
spectrum will yield a temperature measurement).  From Equation~(\ref{eqn:kerv})
this would contribute about 0.019 mag error to the surface brightness 
measurement, which is similar to the effect of a $0.9\%$ error in
$\theta_*$.  Hence, if the current 0.1 mag error in the dereddened
magnitude were not improved, this would be by far the largest error,
equivalent to a 4.6\% error in $\theta_*$.  It is beyond the scope
of the present paper to develop methods to improve this, but I note
that since there have never been any systematic efforts to do so,
it is an open question what might be achieved.

\subsection{{Spectra}
\label{sec:spectra}}

In order to determine whether the color/surface-brightness relation
depends on metallicity, it is of course necessary to measure the
metallicity, which can only be done reliably by taking a spectrum.
For subgiants (also dwarfs), the only cost-effective way to obtain
such spectra is during the microlensing event when the targets
are highly magnified (e.g., \citealt{bensby13} and references therein).
For giant stars, it is practical (albeit more
expensive) to obtain spectra after they have returned to baseline.
Such spectra will automatically yield additional information,
such as the temperature (which can refine the estimate of $(V-K)_0$)
and the radial velocity (to help identify the host population).

\section{{High Metallicity Targets}
\label{sec:targets}}

Of the 56 microlensed dwarfs and subgiants observed by \citet{bensby13},
13 have best fit $\feh>0.3$.  Of these, seven have measured finite
source effects: 
MOA-2008-BLG-311, 
MOA-2008-BLG-310 \citep{mb08310},
MOA-2012-BLG-022, 
OGLE-2007-BLG-349 (\citealt{cohen08}, Dong et al., in prep),
OGLE-2012-BLG-0026 \citep{ob120026},
MOA-2010-BLG-311S \citep{mb10311}, and
MOA-2011-BLG-278 \citep{mb11278}. 
Of these, four have effective temperatures $T_\eff<5500$:
OGLE-2007-BLG-349 (5237),
OGLE-2012-BLG-0026 (4815),
MOA-2010-BLG-311 (5442), and
MOA-2011-BLG-278 (5307).

\begin{table*}[t]
\begin{center}
\caption{\label{tab:events} 
\sc High [Fe/H] Dwarfs and Subgiants with Measured $t_*$}
\vskip 1em
\begin{tabular}{@{\extracolsep{0pt}}lcccc}
\hline
\hline
Event Name   & [Fe/H] & $T_\eff$ & $\mu_\geo$ & $\Delta\theta$(2024) \\ 
             &         &  (K)   & ($\masyr$)& (mas) \\ 
\hline \hline
MOA-2008-BLG-311 & $0.35 \pm 0.08$ & 5947 & 3.7 & 59 \\
MOA-2008-BLG-310 & $0.41 \pm 0.11$ & 5675 & 5.1 & 82 \\
MOA-2012-BLG-022 & $0.42 \pm 0.10$ & 5827 & 1.0 & 12 \\
OGLE-2007-BLG-349 & $0.42\pm 0.26$ & 5237 & 3.1 & 53 \\ 
OGLE-2012-BLG-0026&$0.50\pm 0.44$  & 4815 & 3.7 & 44 \\
MOA-2010-BLG-311S &$0.51 \pm 0.19$ & 5442 & 7.1 & 99 \\
MOA-2011-BLG-278 & $0.52 \pm 0.39$ & 5307 & 4.0 & 52 \\ 
\hline
\hline
\end{tabular}
\end{center}
\end{table*}

I list these with their geocentric proper motions in Table 1, together
with an estimate of their separation (assuming $\mu_\hel = \mu_\geo$) in
2024 when the Giant Magellan Telescope (GMT)
is expected to be fully operational.  The diffraction limit of GMT
at $H$ band is about 17 mas.

I note that there are a number of red giants that have both
$t_*$ measurements and archival spectra.  
\citet{mb9530} report $t_*=2.54\,$days, 
for MACHO-95-30. From 
their reported $K_0=9.83$ and $(V-K)_0=5.03$
I derive $\theta_*=43\,\muas$ and hence $\mu_\geo=6.1\,\masyr$.  They
quote $T_\eff=3700$ but do not attempt to derive a metallicity from their
spectrum.
EROS-2000-BLG-5 has
$\feh=-0.3$, $T_\eff=4500$ \citep{eb20005spec} and $t_*=0.48\,$ days and
$\mu_\geo=5.0\,\masyr$ \citep{eb20005model}.
For OGLE-2002-BLG-069, \citet{ob02069} report $T_\eff=5000$ and $\feh=-0.6$,
while \citet{ob02069model} report
$t_*=0.50\,$days.  I find $\mu_\geo=5.0\,\masyr$ 
from my own notes for this event.
\citet{ob04482} report $T_\eff = 3667$, $t_*=1.25\,$days, $\mu_\geo=7.4\,\masyr$
for OGLE-2004-BLG-482, but do not attempt to estimate a metallicity 
from their spectrum of
this cool $R\sim 40\,R_\odot$ M 
giant\footnote{Note that in text, \citet{ob04482} actually report inconsistent
numbers due to their use of ``$\theta_*$'' for both source radius and
diameter.  The numbers quoted here are the correct ones.}.
\citet{ob04254} report  $\feh=+0.3$, $T_\eff = 4250$, 
$t_*=0.53\,$days, $\mu_\geo=3.1\,\masyr$ for OGLE-2004-BLG-254.

Thus, among these giants, the only one that both has a reported iron
abundance and is metal rich is OGLE-2004-BLG-254.  There are other
giants for which there are microlensed spectra.  For example,
\citet{bensby13} occasionally targetted giants (due to mistakes
in my own photometric source classifications) but did not analyze these for
their sample of ``bulge dwarfs and subgiants''. However, it
is unknown which of these have measured $t_*$ and, in addition
the metallicities are also unknown.

Finally, there are a substantial number of giants that undergo
caustic crossings.  As the new MOA-II and OGLE-IV surveys have
come on line, with many survey observations per night, an
increasing number of these yield measurable $t_*$.  This
number is likely to increase with the advent of the 3-telescope
KMTNet survey in 2015.  As noted above, it is feasible to take
spectra of these giants at baseline (although more convenient
when they are magnified).  Hence, in principle one could
assemble a substantial number with high metallicity for
future measurement of $\mu_\hel$ when they are sufficiently
separated.  

Indeed, \citet{henderson14} has assembled a catalog of 20 microlensing
events with high proper motions $(\ga 8\,\masyr)$ as determined from
finite source effects, including 14 from the literature and six newly
analyzed.  Because of their high-proper motions, many of these may become
suitable targets for proper motion measurements before GMT comes on line.
I find that eight of these are giants and so suitable 
for obtaining post-event (unmagnified) spectra.  These are
MOA-2004-BGL-35 ($8\,\masyr$),
OGLE-2004-BLG-368 ($8\,\masyr$),
OGLE-2004-BLG-482 ($8\,\masyr$),
OGLE-2006-BLG-277 ($13\,\masyr$),
MOA-2007-BLG-146 ($10\,\masyr$),
OGLE-2011-BLG-0417 ($10\,\masyr$),
OGLE-2012-BLG-0456 ($12\,\masyr$), and
MOA-2013-BLG-029 ($9\,\masyr$).  

Note that, as mentioned above,
there is already a spectrum of OGLE-2004-BLG-482 \citep{ob04482},
but no metallicity measurement.  Also note that OGLE-2011-BLG-0417
is one of the very few microlensing events with a complete orbital
solution for the binary lens \citep{ob110417a} and the only that
for which the lens is bright enough to spectroscopically
monitor for radial velocity (RV) varations.  \citet{ob110417} have therefore
advocated an RV campaign in order to test whether the predictions of the
microlensing model are correct.  Since the lens RV is changing by several
$\kms$ from epoch to epoch, while the source is not, 
it should be straight forward to remove
this ``foreground'' and stack the resulting ``decontaminated'' spectra
to obtain a deep spectrum of the source.

Of the 12 non-giant stars in the \citet{henderson14} catalog, 
two were observed by  \citet{bensby13}:  OGLE-2012-BLG-0211 ($\feh=-0.06$)
and MOA-2012-BLG-532 ($\feh=-0.55$).

Even the non-giants that lack spectra can be used to test the overall
method outlined here.  In this regard I note that most of the
\citet{henderson14}
sample have magnified $H$-band data from CTIO-SMARTS.  The exceptions
are 
MOA-2004-BGL-35,
MOA-2011-BGL-040,
OGLE-2012-BLG-0456, and
MOA-2013-BGL-029, which completely lack such data, and
MOA-2011-BGL-262 and
MOA-2011-BGL-274, for which the $H$-band data are of less than top
quality due to low magnification at the time they were taken.

Thus, the prospects are good for applying this technique
to past and future microlensed sources and thereby calibrating the
color/surface-brightness relation at high metallicity.

\acknowledgments

This work was supported by NSF grant AST 1103471 and NASA grant NNX12AB99G.



\end{document}